\documentclass[12pt,amsfonts]{article}
\usepackage{amsfonts}

\def\t{\textstyle}        
\def\one{1\hskip-.37em 1}
\def\ir{{\rm I}\hskip-.2em{\rm R}}
\def\half{{\textstyle{\frac{1}{2}}}}
\def\threebytwo{\textstyle{\frac{3}{2}}}

\def\hg{{\hat g}}
\def\H{{\cal H}}

\def\threebytwo{\textstyle{\frac{3}{2}}}

\def\p{\phi}

\def\H{{\cal H}}

\def\l{\lambda}

\def\S{\Sigma'}
\def\De{\Delta}

\def\tr{{\rm Tr}}
\def\E{{\rm I}\hskip-.2em{\rm E}}
\def\ra{\rightarrow}

\def\tint{{\textstyle\int}}
\def\hg{{\hat g}}
\def\hp{{\hat\pi}}

\def\s{\hskip.08em}

\def\d{\partial}
\def\o{\overline}
\def\a{\alpha}
\def\b{\begin{eqnarray*}}  
\def\e{\end{eqnarray*}}    
\def\bn{\begin{eqnarray}}  
\def\en{\end{eqnarray}}   
\def\<{\langle}
\def\>{\rangle}

\def\no{\nonumber}

\def\k{\kappa}

\def\bx{{\bf x}}

\def\{{\lbrace}
\def\}{\rbrace}

\title{Recent Results Regarding \\ Affine Quantum Gravity}
\author{John R. Klauder\\
Department of Physics and\\Department of Mathematics\\
University of Florida\\
Gainesville, FL 32611-8440}
\date{ }

\begin{document}
\maketitle
\begin{abstract}Recent progress in the quantization of nonrenormalizable scalar fields has found that
a suitable non-classical modification of the ground state wave function leads to a result that eliminates term-by-term divergences that arise in a conventional perturbation analysis. After a brief review of both the scalar field story and the affine quantum gravity program, examination of the procedures used in the latter
surprisingly shows an analogous formulation which already implies that affine quantum gravity
is not plagued by divergences that arise in a standard perturbation study. Additionally, guided by
the projection operator method to deal with quantum constraints, trial reproducing kernels are introduced that satisfy the diffeomorphism constraints. Furthermore, it is argued that the trial reproducing kernels for the diffeomorphism constraints may also satisfy the Hamiltonian constraint as well.
\end{abstract}
\section{Introduction}
Affine quantum gravity is a program for quantization of the Einstein gravitational field in four-dimensional spacetime that differs in many respects from the approach taken in string theory or in loop quantum gravity. The usual problems encountered in quantizing gravity such as: nonrenormalizability, enforcing the gravitational constraints, etc., arise in affine quantum gravity as well, and some of these issues have already been discussed in the two
principal papers on this subject \cite{AQG1,AQG2}. Since their publication, the author has focussed on developing a general understanding of nonrenormalizability as it appears in the simpler case of scalar fields. That focus has led to an alternative---and non-perturbative---approach to such problems that has, in the author's judgement, developed  a viable procedure \cite{IOP,ARX} that is term-by-term divergence free and which restores  nontriviality. In Sec.~2 we present a brief review of how nonrenormalizability in scalar fields is overcome. In Sec.~3 we offer a short summary of the basic principles of affine quantum gravity, and in Sec.~4 we see how the
very solution procedure that works for scalar fields is already built into the procedures used for affine quantum gravity, and as such, typical divergences simply do not appear. This welcome feature is explored in Sec.~5 regarding its general consequences for operator representations and operator product expansions to define local operator products. In Sec.~6  we review the important concept of reproducing kernels and
the reproducing kernel Hilbert spaces they give rise to. In Sec.~7 we give a qualitative assessment of how constraints---initially just the diffeomorphism constraints---can be incorporated into the overall picture, and we also present some sub space examples that fully satisfy diffeomorphism invariance.  In Sec.~8  we discuss the formulation of the affine coherent state overlap function by means of a functional integral. Section 9 is devoted to a discussion of enforcing all of the gravitational constraints including the Hamiltonian constraint, and it is conjectured that the sub spaces designed to fit the diffeomorphism constraints in Sec.~7 already satisfy all of the gravitational constraints. Finally, in Sec.~10 we present
our conclusions.

It is our hope that a continuing study of affine quantum gravity may help illuminate this difficult problem.

\section{Review of Divergence-free Formulation\\ of Scalar Fields}
In \cite{IOP, ARX}, the cause of---and the cure for---divergences in the quantization of covariant scalar fields
was elaborated, and a principal purpose of the present paper is to illustrate similar curative features
already built into the formalism of affine quantum gravity. We begin with a brief overview of the relevant
concepts and results for scalar fields.  Although the general concepts apply to covariant models, as detailed in \cite{IOP}, to keep matters simple here, we consider an idealized (non-covariant, ultralocal) $3+1$ model
with a classical ($c$) Hamiltonian given by
 \bn H_c=\tint\{\half[\pi_c(x)^2+m_0^2\s \p_c(x)^2]+g_0\s\p_c(x)^4\}\,d^3\!x\;, \en
 where $\pi_c(x)$ and $\p_c(x)$, $x\in{\mathbb R}^3$, denote the classical momentum and the classical field. Canonical quantization of the free
 model ($g_0\equiv 0$) leads to a Gaussian ground state as well as an  infinite zero-point energy. To minimize potential ambiguities,
 it is helpful to first regularize the model to eliminate divergences in an effort to determine their cause(s).
    Therefore, let us pass to a hypercubic lattice regularization of ${\mathbb R}^3$ with a lattice
    spacing $a>0$ and $L<\infty$ sites along each edge. We let multi-integers $k=(
    k_1,k_2,k_3)\in {\mathbb Z}^3$ label the sites, and $\pi_k=-i\s a^{-3}\d/\d\p_k$, $\p_k$, $f_k$, etc., denote $q$-numbers and $c$-numbers assigned to each site; here we have set $\hbar=1$.  In that case the lattice-regularized quantum Hamiltonian for the free model becomes
      \bn \H_{latt}=\half\S_k[\pi_k^2+m_0^2\s\p_k^2-m_0\s\hbar\s a^{-3}]\,a^3\;, \en
      where $\S_k$ signifies a sum over the $N'\equiv L^3$ lattice sites.
      The lattice-regularized characteristic function (i.e., Fourier transform) of the ground-state distribution is given by
       \bn C_{latt}(f)\hskip-1.3em&&=M\int e^{i\S_k f_k\s\p_k\,a^3-m_0\s\S_k\p_k^2\,a^3}\,\Pi'_k d\p_k\no\\
          &&= e^{-(1/4m_0)\S_k f_k^2\,a^3 }\no\\
          &&\ra e^{-(1/4m_0)\tint f(x)^2\,d^3x}\;, \en
        where the normalization factor $M=(m_0/\pi)^{N'/2}$, and in the last line we have taken the continuum limit.
      In addition, the lattice version of selected moments in the ground-state distribution are given by
      \bn I_p\equiv M\int [\S_k\s\p_k^2\, a^3]^p\,e^{-m_0\S_k\p_k^2\,a^3}\,\Pi'_k d\p_k=O(N'^{\s p})\;, \en
      where the approximate evaluation arises simply because there are $p$
      such sums and $N'$ terms in each sum, each of which is $O(1)$. In the continuum limit, defined by $a\ra0$ and $L\ra\infty$ such that
      $aL$ is {\it fixed and finite}, all moments for $p\ge1$ diverge; for a discussion of an infinite spatial volume  involving the further limit $aL\ra\infty$, see \cite{IOP}.

      The cause of these divergences is typically attributed to the infinite number of variables, but, in general,
      such integrals need not lead to divergences. Instead, the true cause of such divergences lies in the fact that
      a change of the mass parameter, i.e., $m_0\ra m_0'\,(\not= m_0)$, {\it leads to mutually singular measures as $N'\ra\infty$ that have disjoint support}. For the present example, this property holds because the {\it random} variable $Y_{N'}\equiv N^{'\s-1}\S_k \p_k^2$, becomes, as $N'\ra\infty$ and
       with probability one, {\it deterministic}. Specifically, $\lim_{N'\ra\infty}\s Y_{N'}=1/(2\s m_0)$, a relation that  prescribes the support of such a measure. Change $m_0$ to $m'_0$, and the support changes completely!

       Exposing the source of such divergences becomes immediate if we introduce hyperspherical coordinates
       where $\p_k\equiv\k\s\eta_k$, $\k^2\equiv\S_k\p_k^2$, $1\equiv\S_k\eta_k^2$, with $0\le\k<\infty$ and
       $-1\le\eta_k\le1$. In these variables the expression for the moments becomes
       \bn I_p=M\int [\k^2\,a^3]^p\,e^{-m_0\s\k^2\,a^3}\,\k^{N'-1}\s d\k\,2\s\delta(1-\S_k\eta_k^2)\,\Pi'_k d\eta_k\;. \en
       In this form, divergences arise as $N'\ra\infty$ because a steepest descent evaluation of the $\k$ integral ensures
       that $\k^2=O(N')$, and hence the moment is $O(N^{'\s p})$ as required. To render all such integrals finite, it suffices to arrange that the measure factor $\k^{N'-1}$ is replaced by $\k^{R-1}$ where $R$ is fixed  and finite in order that, as the mass varies, the underlying measures are {\it equivalent} (having equal support) rather than {\it mutually singular} (having disjoint support). We refer to this modification
       of the $\k$-measure factor as {\it measure mashing}.

       To secure this goal, we modify the quantum Hamiltonian by introducing a nonclassical---specifically $O(\hbar^2)$---``counterterm'', which is designed so that the ground-state distribution effectively mashes the $\k$ measure factor from $\k^{N'-1}$ to $\k^{R-1}$. In this effort, we choose $R=2\s b\s a^3N'$, where $b>0$ is a fixed constant with
         dimensions (Length)$^{-3}$ to make $R$ dimensionless. For the model at hand, the resultant quantum Hamiltonian then becomes
         \bn \H'_{latt}=\half\S_k[\pi_k^2+m_0^2\s\p_k^2+\hbar^2\s F\s\p_k^{-2}-E'_0]\,a^3\;, \label{e4}\en
         where $F\equiv (\half-b\s a^3)(\threebytwo-b\s a^3)\,a^{-6}$, and $E'_0=b\s\hbar\s m_0\s a^3$ (which actually vanishes in the continuum limit!) The model given by (\ref{e4}) corresponds to what we have called the {\it pseudofree} theory. To deal with a quartic interaction, it suffices to introduce $g_0\s\S_k\p^4_k\,a^3$ into (\ref{e4}) and change $E'_0$ as needed.

         Specifically, the characteristic function of the ground state distribution for the pseudofree theory is given by
         \bn C'_{latt}(f)\hskip-1.3em&&=M'\int e^{i\s\S_k\s f_k \s\p_k a^3-m_0\S_k\s\p^2_k\s a^3}\,\Pi'_k|\p_k|^{-(1-2ba^3)}\,
         \Pi'_k\s d\p_k\no\\
           &&\ra \exp\{-b\tint d^3x\s\tint[1-\cos(f(x)\s\l)]\,e^{-m\s b\s\l^2}\,d\l/|\l|\}\;, \label{ePOI}\en
           where $m_0=(b\s a^3)\, m$, $\l=\p\s\s a^3$, and in the last line the continuum limit has been taken; it should be noted that the lattice integrand maintains ultralocal symmetry and has the property that its measure has been mashed.  For interacting models the $\l$ dependence of the integrand changes
           but it always retains its dependence as $|\l|^{-1}$ near the origin; all interacting models are
           thus associated with a similar generalized Poisson structure \cite{IOP}.

         Our discussion so far has focused on what happens at a single Euclidean time slice. The implication that finite behavior at a fixed time implies finite behavior in a Euclidean spacetime functional integral formulation follows by means of a straightforward inequality \cite{IOP}, or more directly, by noting that an analogous path integral for a particle has a finite perturbation analysis provided there is a finite perturbation analysis at a sharp time.

         Although some details are different, a rather similar discussion applies to divergences that arise for covariant quantum scalar fields in which, effectively, the same measure mashing
         procedure is used; see \cite{IOP,ARX}.

\section{Basic Affine Quantum Gravity}
The classical canonical formulation of $3+1$ Einstein gravity involves the classical momentum tensor density  $\pi^{ab}(x)\,[\s=\pi^{ba}(x)\s]$ and the spatial metric tensor $g_{ab}(x)\,[\s=g_{ba}(x)\s]$, where $a,b=1,2,3$
\cite{ADM}. Of fundamental importance is {\it metric positivity} which means that $u^a\s g_{ab}(x)\s u^b>0$ for all $x$ and all non-vanishing real vectors $u^a$. Because $\pi^{ab}(x)$ is the generator of translations of $g_{ab}(x)$, this variable may lead to a violation of metric positivity in the classical theory; for example, if $\pi[u]\equiv\tint[u_{ab}(x)\s\pi^{ab}(x)]\s d^3\!x$ is used to generate a macroscopic canonical transformation, then
\bn e^{\{\s\cdot\s, \pi[u]\}}\,g_{ab}(x)\hskip-1.3em&&\equiv g_{ab}(x)+\{g_{ab}(x),\pi[u]\}+\half\{\{g_{ab}(x),\pi[u]\},\pi[u]\}+\cdots\no\\
  && =g_{ab}(x)+u_{ab}(x)\;,\en
where $\{\s\cdot\s,\s\cdot\s\}$ denotes the Poisson bracket with the basic relation
   \bn  \{ g_{ab}(x)\s,\s\pi^{cd}(x')\}=\half(\delta^c_a\s\delta^d_b+\delta^c_b\s\delta^d_a)\s\delta(x,x')\;,\en
 leading to a result that
could violate metric positivity. To eliminate this possibility, and
guided by basic properties of the affine group \cite{AQG1}, we replace the classical momentum tensor $\pi^{ab}(x)$ by the classical ``momentric'' tensor defined by $\pi^a_c(x)\equiv \pi^{ab}(x)\s g_{bc}(x)$, a variable which is so named
to reflect its {\it momen}tum and me{\it tric} ingredients.\footnote{The field $\pi^a_b(x)$, called here the momentric, has also been called the dilation field or the scale field by the author.} Unlike the momentum, the momentric rescales the metric tensor in such a way as to preserve metric positivity; specifically, if $\pi(\gamma)\equiv \tint[ \gamma^a_b(x)\s\pi^b_a(x)\s]\s d^3\!x$ is now used to generate a macroscopic canonical transformation, then
  \bn e^{\{\s\cdot\s, \pi(\gamma)\}}\,g_{ab}(x)\hskip-1.3em&&\equiv g_{ab}(x)+\{g_{ab}(x),\pi(\gamma)\}+\half\{\{g_{ab}(x),\pi(\gamma)\},\pi(\gamma)\}+\cdots\no\\
  && =M^c_a(x)\,g_{cd}(x)\,M^d_b(x)\;,\en
   with $M^c_a(x)\equiv   \{\exp[\gamma(x)/2]\}^c_a$ and $\gamma(x)\equiv\{\gamma^c_a(x)\}$. Thus a metric that satisfies metric positivity will still satisfy metric positivity after such a transformation.

   To preserve metric positivity, it follows that the preferred kinematical variables are $\pi^a_b(x)$ and $g_{ab}(x)$, and it is these
   variables that are promoted to quantum operators in the affine quantum gravity program. {\bf[}{\bf  Remark: }
   The word ``affine'' refers to transformations that map straight lines into straight lines such as
   $x\ra x'=a\s x+b$, $a(\not=0)$, $x$ and $ b\in{\mathbb R}$.
   Our use refers to higher-dimensional spaces for which
   ${\sf x}\ra {\sf A}\s {\sf x}+{\sf b}$ where {\sf x} and {\sf b}$\,\in{\mathbb R}^s$ and ${\sf A}$ is an invertible, $s\times s$ real matrix.{\bf]}

\subsection{Affine commutation relations}
  We let the symbols $\hp^a_b(x)$ and $\hg_{ab}(x)$ denote local self-adjoint operators, which means that they both become self-adjoint operators after smearing with suitable real test functions. Adopting the
  algebra of the classical Poisson brackets for these variables, modified by $i\hbar$ (in units where $\hbar=1$ throughout), the affine commutation relations are given by the formal Lie algebra
\bn  &&[\hp^r_k(x),\,\hp^s_l(y)]=i\,\half\,[\delta^s_k\,\hp^r_l(x)-\delta^r_l\,\hp^s_k(x)]\,\delta(x,y)\;,\no\\
    &&[\hg_{kl}(x),\,\hp^r_s(y)]=i\,\half\,[\delta^r_k\,\hg_{ls}(x)+\delta^r_l\,\hg_{ks}(x)]\,\delta(x,y)\;,\\
  &&[\hg_{kl}(x),\,\hg_{rs}(y)]=0\;. \no \label{e66}\en
  We are interested in representations of these operators, and since, following Dirac \cite{DIR}, quantization should
  precede reduction by any constraints, we deliberately choose an ultralocal representation composed solely of
  independent representations at each spatial point. To regularize such a construction, we again appeal to a
  spatial lattice regularization in which every site carries an independent representation of the affine
  algebra. At a generic lattice site the Lie algebra of the affine group  has the form [with $\hp^a_b\ra \tau^a_b\s \De^{-1}$, where $\De$ denotes a uniform cell volume,  and $\hg_{ab}\ra\sigma_{ab}$] given  by
  \bn  &&\hskip.25cm [\tau^a_b,\tau^c_d]=i\half\s(\delta^c_b\s\tau^a_d-\delta^a_d\s\tau^c_b)\;,
\no\\
  &&\hskip.125cm[\sigma_{ab},\tau^c_d]=i\s\half\s(\delta^c_a\s \sigma_{db}+\delta^c_b\s \sigma_{ad})\;,  \\
  &&[\sigma_{ab},\sigma_{cd}]=0\;. \no \en
Following \cite{ASKL} closely, we choose the faithful, irreducible representation for which the operator matrix $\{\sigma_{ab}\}$ is symmetric and positive definite, and which is unique up to unitary equivalence. Furthermore, we choose a representation which diagonalizes $\{\sigma_{ab}\}$ as $k\equiv\{k_{ab}\}$, $k_{ab}=k_{ba}\in{\mathbb R}$, which we refer to as the $k$-representation; note well: the role of $k$ in this section is very different from the use of that symbol  in Sec.~2. In the associated $L^2$ representation space, and for arbitrary real matrices $\Pi=\{\Pi^{ab}\}$, $\Pi^{ba}=\Pi^{ab}$, and $\Gamma=\{\Gamma^c_d\}$, it follows that
 \bn  U[\Pi,\Gamma]\s\psi(k)\hskip-1.3em&&\equiv e^{i\s\Pi^{ab}\s\sigma_{ab}}\,e^{-i\s\Gamma^c_d\s\tau^d_c}\s\psi(k)\no\\
    &&= (\det[S])^2\,e^{i\s\Pi^{ab}\s k_{ab}}\s\psi(SkS^T)\;,  \en
where $S\equiv e^{-\Gamma/2}=\{S^a_b\}$ and $(SkS^T)_{ab}\equiv S_a^ck_{cd}S^d_b$. The given transformation is unitary within the inner product defined by
 \bn \int_+\psi(k)^*\s\psi(k)\,dk \;, \en
where $dk\equiv\Pi_{a\le b}\s dk_{ab}$, and the ``$+$'' sign denotes an integration over only that part of the six-dimensional $k$-space where the elements form a symmetric, positive-definite matrix, $\{k_{ab}\}>0$.
To define affine coherent states we first choose the fiducial vector as an extremal weight vector \cite{ASKL,AQG2},
  \bn  \eta(k)\equiv C\s(\det[k])^{\beta-1}\,e^{-\beta\s\tr[{\tilde G}^{-1}k]}\;,  \en
where $\beta>0$, ${\tilde G}=\{{\tilde G}_{ab}\}$ is a fixed positive-definite matrix, $C$ is determined by normalization, and Tr denotes the trace.
This choice of $\eta(k)=\<k|\eta\>$ leads to the expectation values
 \bn   \<\eta|\s\sigma_{ab}\s|\eta\>\hskip-1.3em&&=\int_+\eta(k)^*\s k_{ab}\,\eta(k)\,dk={\tilde G}_{ab}\;,  \\
    \<\eta|\s\tau^c_d\s|\eta\>\hskip-1.3em&&=\int_+\eta(k)^*\s{\tilde\tau}^c_d\s\eta(k)\,dk=0\;,\no\\
        {\tilde\tau}^c_d\hskip-1.3em&&\equiv (-i/2)[\s(\d/\d k_{cb})\s k_{db}+k_{db}\s(\d/\d k_{cb})\s] \;.\en
In the $k$-representation, it follows that the affine coherent states are given by
  \bn  \<k|\Pi,\Gamma\>\equiv C\s(\det[S])^2\s(\det[SkS^T])^{\beta-1}\,e^{i\s\tr[\Pi\s  k]}\,e^{-\beta\s\tr[{\tilde G}^{-1}SkS^T]} \;.\label{e77}  \en

Observe that what really enters the functional argument above is the positive-definite matrix $G^{-1}\equiv S^T{\tilde G}^{-1}\s S$, where we set $G\equiv\{G_{ab}\}$. Thus, without loss of generality, we can drop the label $\Gamma$ (or equivalently $S$) and replace it with $G$. Hence, the affine coherent states become
  \bn  \<k|\Pi,G\>\equiv C'\s(\det[G^{-1}])^{\beta}\s(\det[k])^{\beta-1}\,e^{i\s\tr[\Pi\s k]}\,e^{-\beta\s\tr[G^{-1}k]}\;,  \en
where $C'$ is a new normalization constant.
It is now straightforward to determine the affine coherent state overlap function
  \bn   &&\hskip-1.5cm\<\Pi'',G''|\Pi',G'\>=\int_+\<\Pi'',G''|k\>\s\<k|\Pi',G'\>\,dk \no\\
  &&  \hskip1.31cm=\bigg[\,\frac{\{\det[G''^{-1}]\det[G'^{-1}]\}^{1/2}}{\det\{\half[(G''^{-1}+G'^{-1})
  +i\beta^{-1}(\Pi''-\Pi')]\}}\,\bigg]^{2\beta} \;.  \label{h30}\en
In arriving at this result, we have used normalization of the coherent states to eliminate the constant $C'$.

Suppose we now consider a $3$-dimensional spatial lattice of independent sets of matrix degrees of freedom and build the corresponding affine coherent state overlap as the product of expressions like (\ref{h30}). In this section, we adopt the notation where $\bf n$ labels a lattice site and ${\bf n}\in{\bf N}$, the set of all spatial lattice sites, which in turn may be considered a finite subset of ${\mathbb Z}^3$; this change of notation conforms with the analysis in \cite{AQG1} and is needed because $k$ has a different meaning in the present section from its use in Sec.~2.
With these notational changes, the multi-site, affine coherent state overlap function is given by
 \bn &&\<\Pi'',G''|\Pi',G'\>_{\bf N} \no\\
&&\hskip1cm=\prod_{{\bf n}\in{\bf N}}\bigg[\,\frac{\{\det[G''^{-1}_{[{\bf n}]}]\det[G'^{-1}_{[{\bf n}]}]\}^{1/2}}{\det\{\half[(G''^{-1}_{[{\bf n}]}+G'^{-1}_{[{\bf n}]})+i\beta^{-1}_{[{\bf n}]}(\Pi''_{[{\bf n}]}-\Pi'_{[{\bf n}]})]\}}\,\bigg]^{2\beta_{[{\bf n}]}} \;;  \label{i24}\en
since there is generally no translation symmetry, it is not necessary that the factors $\beta_{[{\bf n}]}$ are all the same.

\subsection{Affine coherent state overlap function}
As our next step we take a limit in which the number of lattice sites with independent matrix degrees of freedom tends to infinity, but also the lattice cell volume tends to zero so that, loosely speaking, the lattice points approach the continuum points of the underlying topological space $\cal S$. In order for this continuum limit to be meaningful, it is necessary that the exponent $\beta_{[{\bf n}]}\ra0$ in a suitable way. Again assuming a uniform cell volume for simplicity, we set
 \bn  \beta_{[{\bf n}]}\equiv b_{[{\bf n}]}\s\De  \;,  \en
where the cell volume $\De$ has the dimensions (Length)$^3$, and thus $b_{[{\bf n}]}$ has the dimensions (Length)$^{-3}$. In addition, we rename $\Pi^{ab}_{[{\bf n}]}\equiv \pi^{ab}_{[{\bf n}]}\s\De$, as well as $G_{ab\s{[{\bf n}]}} \equiv g_{ab\s{[{\bf n}]}}$, and denote the matrix elements of $G^{-1}_{[{\bf n}]}$ by $g^{ab}_{[{\bf n}]}$. With these notational changes
(\ref{i24}) becomes
\bn &&\<\pi'',g''|\pi',g'\>_{\bf N} \no\\
&&\hskip1cm\equiv\prod_{{\bf n}\in{\bf N}}\bigg[\,\frac{\{\det[g''^{ab}_{[{\bf n}]}]\det[g'^{ab}_{[{\bf n}]}]\}^{1/2}}{\det\{\half[(g''^{ab}_{[{\bf n}]}+g'^{ab}_{[{\bf n}]})+i\s b^{-1}_{[{\bf n}]}(\pi''^{ab}_{[{\bf n}]}-\pi'^{ab}_{[{\bf n}]})]\}}\,\bigg]^{2\s b_{[{\bf n}]}\s\De} \;.  \label{i25}\en
To facilitate the continuum limit, we assume that the various label sets pass to smooth functions, in which
case the result, for a compact topological space ${\cal S}$, is given by
 \bn  &&\hskip-.8cm\<\pi'',g''|\pi',g'\> \no\\
 &&\hskip-.3cm \equiv\exp\bigg[-2\int b(x)\,d^3\!x\,\no\\
&&\hskip.1cm\times\ln\bigg(\frac{\det\{\half[g''^{ab}(x)+g'^{ab}(x)]+\half\s i\s b(x)^{-1}\s[\pi''^{ab}(x)-\pi'^{ab}(x)]\}}{\{\det[g''^{ab}(x)]\,\det[g'^{ab}(x)]\}^{1/2}}\bigg)\bigg].  \label{i22}\en
In this way we see how the continuum result may be obtained as a limit starting from a collection of independent affine degrees of freedom. The necessity of ending with an integral over the space $\cal S$ has directly led to the requirement that we introduce the scalar density function $b(x)>0$, but, other than smoothness and positivity,  this
function may be freely chosen. Indeed, the appearance of $b(x)$ in the affine coherent state overlap function is not unlike the appearance of a Gaussian width parameter (often  denoted by
$\omega$) in the overlap function of canonical coherent states; it is a representation artifact with
limited physical significance.  At this point it may be useful to compare (\ref{i22}) with (\ref{ePOI})
for their similarities and differences.

It is important to appreciate that the function $g_{ab}(x)$ that enters the affine coherent state overlap function is {\it not} a true metric that has been applied to the topological space ${\cal S}$, but it is just a symmetric, two-index covariant tensor field, which we sometime call a metric because of its positivity properties. In like manner, the momentum  field $\pi^{ab}(x)$ that enters the affine coherent state overlap function also carries no special physical meaning and is just a symmetric, two-index  contravariant tensor  density field---nothing more and nothing less.

The expression (\ref{i22}) is not yet in optimal form, especially if one wishes to consider an
extension to an infinite spatial volume.
In this case, it is useful to incorporate the proper metric and momentum asymptotics for large spatial distances. Assuming that the matrix elements $\gamma^a_b(x)$  and $\pi^{ab}(x)$ are smooth functions with compact support, it follows  that for sufficiently large $x$ values, the affine coherent state matrix elements
tend to become $\<\pi'',g''|\s \hg_{ab}(x)\s|\pi',g'\>={\tilde g}_{ab}(x)$ and $\<\pi'',g''|\s\hp^a_b(x)\s|\pi',g'\>= 0$. We can already make use of these facts by freely multiplying the numerator and denominator
inside the logarithm in (\ref{i22}) by $\det[{\tilde g}_{ab}(x)]$, which leads to the relation
\bn  &&\hskip-.8cm\<\pi'',g''|\pi',g'\> \no\\
 &&\hskip-.3cm =\exp\bigg[-2\int b(x)\,d^3\!x\,\no\\
&&\hskip.1cm\times\ln\bigg(\frac{\det\{\half[{\tilde g}''^{a}_b(x)+{\tilde g}'^a_b(x)]+\half i\s b(x)^{-1}[{\tilde \pi}''^a_b(x)-{\tilde\pi}'^a_b(x)]\}}{\{\det[{\tilde g}''^a_b(x)]\,\det[{\tilde g}'^a_b(x)]\}^{1/2}}\bigg)\bigg]\;.  \label{k22}\en
In this equation we have introduced ${\tilde g}^a_b(x)\equiv {\tilde g}_{bc}(x)\s g^{ac}(x)$ and
${\tilde \pi}^a_b(x)\equiv {\tilde g}_{bc}(x)\s \pi^{ac}(x)$ for both the $''$ and the $'$ variables. With this
formulation, it follows that at large distances ${\tilde g}^a_b(x)\ra \delta^a_b$ while ${\tilde\pi}^a_b(x)\ra0$.
Moreover, we note that each one of the three determinants in (\ref{k22}), e.g., $\det[{\tilde g}''^a_b(x)]$, etc., is {\it invariant under general coordinate transformations}, and each determinant tends toward unity at large distances. Thanks to the logarithm, this tendency ensures convergence of the contribution from each determinant in an infinite volume.

Clearly, the case of an infinite spatial volume needs some special care. To focus on the essentials, we now return
to our general assumption that we deal with a {\it finite spatial volume}, unless noted otherwise.

\section{Measure Mashing in the Affine \\Coherent State Overlap}
Next, let us derive the coherent state overlap (\ref{i22}) in a different manner. In particular, with $\{C''_{\bf n}\}$ a presently unimportant set of constants for these integrals, it is clear that
  \bn  &&\hskip-1.5em\<\pi'',g''|\pi',g'\> \no\\
  &&\hskip-.8em=\lim_{\De\ra0} \int_+ \Pi'_{\bf n}\,\<\pi''_{\bf n},g''_{\bf n}|k_{\bf n}\>\<k_{\bf n}|\pi'_{\bf n},g'_{\bf n}\> \,dk_{\bf n}\no\\
  &&\hskip-.8em= \lim_{\De\ra0}\int_+ \Pi'_{\bf n}\,C''_{\bf n}\,e^{-i\s \tr[(\pi''_{\bf n}-\pi'_{\bf n})\s k_{\bf n}]-\beta_{\bf n}\s \tr[(g^{''\s-1}_{\bf n}+g^{'\s-1}_{\bf n})\s k_{\bf n}]} \,\det(k_{\bf n})^{2(\beta_{\bf n}-1)}\,dk_{\bf n}\; ,\no\\ \en
which, on transforming to hyperspherical coordinates, defined here by $k_{{\bf n}\,ab}\equiv \k\s\eta_{{\bf n}\,ab}$, $\k^2\equiv
\S_{\bf n}\Sigma_{a\le b}\,k_{{\bf n}\,ab}^2$, $1\equiv\S_{\bf n}\Sigma_{a\le b}\,\eta_{{\bf n}\,ab}^2$, with $0\le \k<\infty$ and
$-1\le \eta_{{\bf n}\,ab}\le1$, {\it subject as well to the requirement that the matrix $\{\eta_{{\bf n}\,ab}\}$ is positive definite for each ${\bf n}$}, leads  to
   \bn  &&\hskip-.8cm\<\pi'',g''|\pi',g'\> \no\\
    &&\hskip-.2cm= \lim_{\De\ra0}\int_+ \{\Pi'_{\bf n}\,C''_{\bf n}\,e^{-i\s \tr[(\pi''_{\bf n}-\pi'_{\bf n})\s \k\eta_{\bf n}]-\beta_{\bf n}\s \tr[(g^{''\s-1}_{\bf n}+g^{'\s-1}_{\bf n})\s \k\eta_{\bf n}]} \,\det(\k\s\eta_{\bf n})^{2(\beta_{\bf n}-1)}\}\no\\
    &&\hskip3em\times\,\k^{(6\s|\bf N|-1)}\,d\k\;2\s\delta(1-\S_{\bf n}\Sigma_{a\le b}\eta_{{\bf n}\,ab}^2)\Pi'_{\bf n}\,d\eta_{\bf n} \;,\en
    where $|{\bf N}|<\infty$ denotes the total number of lattice sites in this spatial slice.
    With $\beta_{\bf n}= b_{\bf n}\s\De$, it follows  that
       \bn [\Pi'_{\bf n}\det(\k\s\eta_{\bf n})^{2(b_{\bf n}\s\De-1)}]\,\k^{6\s|{\bf N}|-1}=\Pi'_{\bf n}\,[\s \det(\eta_{\bf n})^{2(b_{\bf n}\s\De-1)}\,\k^{6\s b_{\bf n}\s\De}\s]\,\k^{-1}\;, \en
       {\it which shows the effects of measure mashing already} for which, in the present case,  $R\equiv 6\s\S_{\bf n}b_{\bf n}\,\De<\infty$ provided, as we have assumed, that $\S_{\bf n}\,\De= V'<\infty$.

       \subsection{Significance of already having a mashed measure}
       According to the discussion in Sec.~2, a mashed measure for the hyperspherical radius variable
       implies no divergences arise in any perturbation analysis because potentially disjoint measures have been converted into equivalent measures. It also implies that local field powers are defined by means of an operator product expansion and not by normal ordering, as we make clear in the following section.  When dealing with quantum field theories, we normally encounter unitarily inequivalent representations of kinematical operators when a parameter is changed in the basic model. For example, in the idealized free model discussed in Sec.~2, and for a finite spatial volume as well,  representations of the local operators for the momentum $\pi(x)$ and the field $\p(x)$  are unitarily inequivalent for different mass values.
       On the other hand, for a finite spatial volume, the pseudofree model local field operators are unitarily equivalent for different mass values.

In the present case, we are dealing with the affine gravitational field operators $\hp^a_b(x)$ and $\hg_{ab}(x)$, for $a,b=1,2,3$. A change in the scalar density $b(x)$, which enters the representation of the coherent states, could lead
to inequivalent representations of the affine kinematical variables, but---{\it thanks to measure mashing}---that is not the case since all
the representations covered by the affine coherent states are  unitarily equivalent for different functions
$b(x)$. This has the additional consequence that we can construct various operators from the basic kinematical set and these operators will also be unitarily equivalent for different $b(x)$ values as well.

\section{Operator Realization}
In order to realize the metric and momentric fields as quantum operators in a Hilbert space, $\frak H$, it is expedient---following the procedures for ultralocal scalar fields \cite{kla99,kla66}---to introduce a set of conventional local {\it annihilation and creation operators},  $A(x,k)$ and  $A(x,k)^\dag$, respectively, with the only nonvanishing commutator given by
  \bn  [A(x,k),\,A(x',k')^\dag]=\delta(x,x')\,\delta(k,k')\one\;, \en
where $\one$ denotes the unit operator.
Here, $x\in\ir^3$, while $k\equiv\{k_{rs}\}$ denotes a positive-definite, $3\times3$ symmetric matrix degree of freedom confined to the domain where $\{k_{rs}\}>0$; additionally, $\delta(k,k')\equiv \Pi_{a\le b}\s\delta(k_{ab}-k'_{ab})$. We introduce a ``no-particle'' state $|0\>$ such that $A(x,k)\,|0\>=0$ for all arguments. Additional states
are determined by suitably smeared linear combinations of
  \bn A(x_1,k_1)^\dag\,A(x_2,k_2)^\dag\,\cdots\,A(x_p,k_p)^\dag\,|0\> \en
for all $p\ge1$, and the linear span of all such states is $\frak H$ provided,
apart from constant multiples, that $|0\>$ is the only state annihilated by all the $A$ operators. Thus we are led to a conventional Fock representation for the $A$ and $A^\dag$ operators. Note that the Fock operators are irreducible, and thus all operators acting in $\frak H$ are given as suitable functions of them.

Next, let $c(x,k)$ be a possibly complex, $c$-number function (defined below) and introduce the translated Fock operators
  \bn  B(x,k)\hskip-1.4em&&\equiv A(x,k)+c(x,k)\,\one\;,\no \\
     B(x,k)^\dag\hskip-1.4em&& \equiv A(x.k)^\dag + c(x,k)^*\,\one\;. \en
Evidently, the only nonvanishing commutator of the $B$ and $B^\dag$ operators is
  \bn  [B(x,k),\,B(x',k')^\dag]=\delta(x,x')\,\delta(k,k')\one\;, \en
the same as the $A$ and $A^\dag$ operators. With regard to transformations of the coordinate $x$, it is clear that $c(x,k)$ (just like the local operators $A$ and  $B$) should transform as a scalar density of weight one-half. Thus we set
  \bn c(x,k)\equiv b(x)^{1/2}\,d(x,k)\;, \en
where $d(x,k)$ transforms as a scalar. The criteria for acceptable $d(x,k)$ are, for each $x$, that
  \bn &&\tint_+\,|d(x,k)|^2\,dk=\infty\;,\label{pp2} \\
&&\tint_+\,k_{rs}\,|d(x,k)|^2\,dk = \delta_{rs}\;,\label{pp3}  \en
the latter assuming that ${\tilde g}_{kl}(x)=\delta_{kl}$.     

We shall focus only on the case where
  \bn d(x,k)\equiv \frac{ \sqrt{K}\,e^{-\tr (k)}  }{\det (k)}\;, \en
which is everywhere independent of $x$; $K$ denotes a positive constant to be fixed by (\ref{pp3}). The given choice for $d$ corresponds to the case where ${\tilde\pi}^{kl}(x)\equiv0$ and ${\tilde g}_{kl}(x)\equiv\delta_{kl}$. {\bf[Remark:} For different choices of asymptotic fields it suffices to choose
  \bn d(x,k)\ra {\tilde d}(x,k)\equiv \frac{ \sqrt{K}\,e^{ -i\s b(x)^{-1}{\tilde\pi}^{ab}(x)\s k_{ab} }\,e^{ -{\tilde g}^{ab}(x)\s k_{ab} }  }{\det (k)}\;.  \en
It would seem that such cases allow one to quantize about a variety of topological backgrounds.{\bf]}

In terms of the quantities introduced above, the local metric operator is defined by
 \bn \hg_{ab}(x)\equiv b(x)^{-1}\tint_+B(x,k)^\dag\,k_{ab}\,B(x,k)\,dk\;,\en
and the local momentric operator is defined by
 \bn \hp^r_s(x)\equiv -i\half\tint_+B(x,k)^\dag\,(k_{st}\d^{tr}+\d^{rt}k_{ts})\,B(x,k)\,dk\;.\en
Here $\d^{st}\equiv\d/\d k_{st}$, and $\hg_{ab}(x)$ transforms as a tensor while $\hp^r_s(x)$ transforms as a tensor density of weight one.
It is straightforward to show that these operators satisfy the required affine commutation relations, and moreover that \cite{kla99,kla66}
  \bn&& \hskip.0cm\<0|\,e^{i\tint\pi^{ab}(x)\hg_{ab}(x)\,d^3\!x}\,
e^{-i\tint\gamma^s_r(x)\hp^r_s(x)\,d^3\!x}\,|0\> \no\\
&&\hskip-.7cm=\exp\{-K\tint b(x)\,d^3\!x\tint[e^{-2\delta^{ab}k_{ab}}-e^{-i\pi^{ab}(x)k_{ab}/b(x)}e^{-[(\delta^{ab}+g^{ab}(x))k_{ab}]}]\,dk/(\det k)^2\} \no\\
&&\hskip-.7cm =\exp[\!\![-2\tint b(x)\,d^3\!x\ln(\!\!([\det(g_{ab})]^{1/2}\,\det\{\half[\delta^{ab}+g^{ab}(x)]-i\half b(x)^{-1}\pi^{ab}(x)\})\!\!)\,]\!\!]\;,
\no\\  &&\en
where $K$ has been chosen so that 
  \bn K\tint_+ k_{rs}\,e^{-2\,\tr(k)}\,dk/(\det k)^2=\delta_{rs}\;.  \en
An obvious  extension of this calculation leads to (\ref{i22}).

\subsection{Local operator products}
Basically, local products for the gravitational field operators follow the pattern for other ultralocal quantum field theories \cite{kla99,kla66}. As motivation, consider the product
  \bn  &&\hskip-.7cm\hg_{ab}(x)\hg_{cd}(x')\no\\
&&=[b(x)\s b(x')]^{-1}\tint_+B(x,k)^\dag\,k_{ab}\,B(x,k)\,dk\,\cdot\, \tint_+B(x',k')^\dag\,k'_{cd}\,B(x',k')\,dk'\no\\
&&=[b(x)\s b(x')]^{-1}\tint_+\tint_+B(x,k)^\dag\,k_{ab}\,[\,B(x,k),\,B(x',k')^\dag]k'_{cd}\,B(x',k')\,dk\,dk'\no\\
&&\hskip2cm+!\s\hg_{ab}(x)\hg_{cd}(x')\s!\no\\
&&=b(x)^{-2}\delta(x,x')\tint_+B(x,k)^\dag k_{ab}\,k_{cd}\,B(x,k)\,dk+!\s\hg_{ab}(x)\hg_{cd}(x')\s!\;, \en
where $!\; !$ denotes normal ordering with respect to $B^\dag$ and $B$.
When $x'\ra x$, this relation formally becomes
\bn\hg_{ab}(x)\hg_{cd}(x)=b(x)^{-2}\delta(x,x)\tint_+B(x,k)^\dag k_{ab}\,k_{cd}\,B(x,k)\,dk+!\s\hg_{ab}(x)\hg_{cd}(x)\s! \en
After formally dividing both sides by the most divergent dimensionless scalar factor, namely, the ``scalar'' $b(x)^{-1}\delta(x,x)$,
we define the renormalized (subscript ``$R$'') local product as
  \bn [\hg_{ab}(x)\hg_{cd}(x)]_R\equiv b(x)^{-1}\tint_+B(x,k)^\dag\,k_{ab}\,k_{cd}\,B(x,k)\,dk \;;\en
  note: this derivation can be made rigorous with test functions and proper limits.
   Observe that $[\hg_{ab}(x)\hg_{cd}(x)]_R$  is a local operator that becomes a self-adjoint operator when smeared with a suitable real test function just as much as the local operator $\hg_{ab}(x)$ becomes a self-adjoint operator after smearing with a suitable real test function. Higher-order local operator products exist as well, for example,
  \bn  &&[\hg_{a_1b_1}(x)\hg^{a_2b_2}(x)\hg_{a_3b_3}(x)\cdots\hg_{a_pb_p}(x)]_R\no\\
&&\hskip1.5cm\equiv b(x)^{-1}\tint_+B(x,k)^\dag\,(k_{a_1b_1}k^{a_2b_2}k_{a_3b_3}\cdots k_{a_pb_p})\,B(x,k)\,dk\,, \en
which, after contracting on $b_1$ and $b_2$, implies that
\bn [\hg_{a_1b}(x)\hg^{a_2b}(x)\hg_{a_3b_3}(x)\cdots\hg_{a_pb_p}(x)]_R=\delta^{a_2}_{a_1}\,[ \hg_{a_3b_3}(x)\cdots\hg_{a_pb_p}(x)]_R \;.\en
It is in this sense that $[\hg_{ab}(x)\hg^{bc}(x)]_R=\delta^c_a\s\one$.

Diagonal coherent state matrix elements are of particular interest. As an example, consider
  \bn \<\pi,g|\s \hg_{ab}(x)\s|\pi,g\>\hskip-1.4em&&= \<0,g|\s\hg_{ab}(x)\s|0,g\>\no\\
      &&= M(x)^c_a\,\<k_{cd}\>\,M^d_b(x)\equiv g_{ab}(x)\;, \en
which determines the meaning of the label function $g_{ab}(x)$,  and where we have set 
   \bn \< (\,\cdot\,) \s\>\equiv K\tint_+ (\,\cdot\,)\, e^{-2\tr(k)}\,dk/\det( k)^2\;. \en
Likewise,
    \bn  &&\hskip-2em\<\pi,g|\s [\hg_{ab}(x)\s\hg_{ef}(x)\s]_R\s|\pi,g\>\no\\
    &&= \<0,g|\s [\hg_{ab}(x)\s\hg_{ef}(x)\s]_R\s|0,g\>\no\\
      &&= M(x)^c_a\,M^g_e(x)\,\<k_{cd}\s k_{gh}\>\,M^d_b(x)\, M^h_f(x) \no\\
      &&= g_{ab}(x)\s g_{ef}(x) \no\\
      &&\hskip2em + M(x)^c_a\,M^g_e(x)\,\<[k_{cd}\s k_{gh}-\<k_{cd}\>\<k_{gh}\>]\>\,M^d_b(x)\, M^h_f(x)\;.
      \label{p34}\en

      The latter expression brings up a question dealing with the classical limit in which $\hbar\ra0$. So far we have kept $\hbar=1$, and its true role in the story is not evident.
      The important factor in (\ref{p34}) given by
         \bn \<[k_{cd}\s k_{gh}-\<k_{cd}\>\<k_{gh}\>]\>\equiv \<k_{cd}\s k_{gh}\>^T \en
         should be $O(\hbar)$ so that in the classical limit we would find that
         \bn \lim_{\hbar\ra0}\,\<\pi,g|\s [\hg_{ab}(x)\s\hg_{ef}(x)\s]_R\s|\pi,g\>=g_{ab}(x)\s g_{ef}(x)\;.\label{p35}\en
         This dependence can be arranged as follows. Restoring $\hbar$ for the moment,
         we propose that (\ref{e77}), the elementary building block of an affine coherent state at a single site in a lattice regularization, is given instead  by
         \bn \<k|\Pi,\Gamma\>=C\s(\det[S])^2(\det[SkS^T])^{({\tilde\beta}/\hbar)-1}\,e^{(i/\hbar)\s\tr(\Pi\s k)-({\tilde\beta}/\hbar)\s\tr({\tilde G}^{-1}S\s k\s S^T)}\;, \en
              where ${\tilde\beta}\equiv \beta\s\hbar$ but what is different now is that we treat $\tilde\beta$ and $\hbar$ as two {\it independent} variables and when $\hbar\ra0$, then $\tilde\beta$ remains fixed.
           This has the desired effect of making the truncated expression $\<k_{cd}\s k_{gh}\>^T\propto\hbar\ra0$,  as $\hbar\ra0$, leading to (\ref{p35}).

           A similar story holds for the momentric operator.

\subsection{Advantages of the operator product expansion}
The implications of the operator product analysis given above are quite significant
in defining the Hamiltonian constraint operator.
Ignoring its constraint nature, the classical expression is given by
  \bn H(x)= g(x)^{-1/2} \,[\s\pi^a_b(x)\s\pi^b_a(x)-\half\s\pi^a_a(x)\s\pi^b_b(x)\s]+g(x)^{1/2}\, R(x)\;.\en
  Quantization of this expression may be interpreted as
  \bn \H(x)\hskip-1.4em&&= [\s\hp^a_b(x)\s \hg(x)^{-1/2}\s\hp^b_a(x)]_R-\half\s[\hp^a_a(x)\s \hg(x)^{-1/2}\s\hp^b_b(x)\s]_R \no\\
  &&\hskip4em +[\hg(x)^{1/2}\,{\hat R}(x)\s]_R\;, \en
  where each of the terms is a local operator such that when integrated over a finite spatial region
  each term, as well as the whole expression, leads to a self-adjoint operator.
  This simple
  conceptual analysis arises because local operator products are derived from operator product expansions.

      \section{Properties and Virtues of \\Reproducing Kernel Hilbert Spaces}
      We have focussed on coherent states and properties of the coherent state overlap function
      for a very good reason: such states provide an unusually convenient bridge between the classical
      and quantum theories. Traditionally, coherent states are associated with a local resolution of unity, but the affine coherent states that are of interest to us do  not possess a local resolution of unity;
      instead, they constitute what we have called a family of {\it weak coherent states} \cite{GUTZ}. To generate a representation of the kinematic Hilbert space of interest without invoking a local integral resolution of unity,
      we need to appeal to another formulation. The manner in which this is done is quite general and it is convenient to discuss this procedure initially in general terms not limited to the affine coherent states that are the principal focus of our study.

      Consider an $L$-dimensional label space ${\cal L}$, composed of points $l\in{\cal L}$, that form a topological space; in other words, the space ${\cal L}$ admits a notion of continuity, $l'\ra l$, in the sense of ${\cal L}$. When $L<\infty$, it generally suffices to assume that the label $l=(l_1,l_2,\ldots,l_L)$, with $l_j\in{\mathbb R}$, and that ${\cal L}$  is locally isomorphic to an $L$-dimensional Euclidean space ${\mathbb R}^L$ so that convergence is simply convergence of each coordinate. Infinite dimensional labels spaces are normally natural generalizations of finite ones.

      Next, consider a function ${\cal K}(l'';l')$ that is {\it jointly continuous in both labels}, and which satisfies the condition
        \bn \Sigma_{i,j=1}^{N,N} \a^*_i\s \a_j\,{\cal K}(l_i;l_j)\ge0 \en
        for all complex coefficients $\{\a_j\}$ and all $N<\infty$. {\it Such a function is said to be a continuous function of
        positive type}, and it will be the foundation of a very useful representation space \cite{MESH}. According to the GNS (Gel'fand, Naimark, Segal) Theorem \cite{EMCH}, there is always an abstract, separable Hilbert space with
        vectors $|l\>$ labeled by points in $l\in{\cal L}$ such that ${\cal K}(l'';l')\equiv \<l''|l'\>$.
        The function ${\cal K}(l'';l')$ is called a {\it reproducing kernel} (for reasons to be made clear below). As the notation suggests, the affine coherent state overlap function is a function of positive type and therefore qualifies as a reproducing kernel.

        To generate an appropriate Hilbert space representation, we envisage two abstract vectors as members
        of a dense set of vectors given by
        \bn   |\psi\>\hskip-1.3em&&\equiv{\t\sum}_{i=1}^I\s\a_i\,|l_{(i)}\>\;, \no\\
              |\p\>\hskip-1.3em&&\equiv{\t\sum}_{j=1}^J \beta_j\,|l_{[j]}\>\;, \label{e35}\en
              where $I$ and $J$ are both finite.
      To give a functional representation of these abstract vectors, we appeal to the only set of vectors that we know span the Hilbert space, and that is the set $\{|l\>\}$. Thus we introduce
        \bn   \psi(l)\hskip-1.3em&&\equiv {\t\sum}_{i=1}^I\s\a_i\,\<l|l_{(i)}\> =\<l|\psi\>\;,\no\\
              \p(l)\hskip-1.3em&&\equiv {\t\sum}_{j=1}^J \beta_j\,\<l|l_{[j]}\>=\<l|\p\>\;, \en
            which yield continuous functional representatives. To define the inner product of two such vectors we appeal to (\ref{e35}) and let
            \bn (\psi,\p)\equiv {\t\sum}_{i,j=1}^{I,J}\,\a^*_i\s  \beta_j\,\<l_{(i)}|l_{[j]}\>=\<\psi|\p\>\;. \en
            The name ``reproducing kernel'' arises because if $\a_{i}=\delta_{1,i}$, then
               \bn (\<\cdot|l'\>,\p)={\t\sum}_{j=1}^J \beta_j\,\<l'|l_{[j]}\>=\p(l')\;;\en
               in short, the element $\p(l)$ is reproduced by this inner product.

            All that remains to generate a functional Hilbert space is to complete the space by including the limits of all Cauchy sequences. The result is a Hilbert space representation ${\cal C}$ composed entirely of continuous functions on ${\cal L}$; note there are no sets of measure zero in this approach: every vector is represented by a unique, continuous function. Observe that every aspect of the Hilbert space ${\cal C}$  is determined by the reproducing kernel!

            Traditional coherent states also generate a reproducing kernel Hilbert space which has the property that it has two independent rules for evaluating inner products: the first by a local integral with a non-negative, absolutely continuous measure, and the second in the manner just described for a reproducing kernel Hilbert space. However, not all reproducing kernel Hilbert spaces involve local integral inner products, and the affine coherent state overlap function
               is of that type  leading to a continuous function of positive type but not one with a local integral resolution of unity. As already noted, when that happens, we say that we are dealing with weak coherent states. This is  already true even for just a single site whenever $0<\beta\le1/2$; to see this, try integrating the absolute square of (\ref{h30}) over the group-invariant measure.

\subsection{Reduction of a reproducing kernel}
         It sometimes happens that the space spanned by the set of states $\{|l\>\}$ is the same as spanned by a suitable subset of that set, as for example, the set given by $\{|l^*\>\}$, where
         $|l^*\>\equiv\tint\,|l\>\,\sigma(l)\,dl$, and  $\sigma(l)\,dl$ denotes a possibly complex, fixed measure. Stated otherwise, if we denote the linear span of vectors by an overbar, then  $\overline{|l^*\>}\equiv\overline{\tint\,|l\>\,\sigma(l)\,dl}=\overline{|l\>}$. A common situation arises when the measure $\sigma\,dl$ is a delta measure fixing one or more of the coordinates. We next discuss a few concrete examples.

         To make this discussion more relevant to our general discussion, let us focus on the reproducing kernel given by the affine coherent state overlap function $\<\pi'',g''|\pi',g'\>$ as defined in (\ref{k22}). Two forms of reduction of this reproducing kernel are worth discussing.

         In the first form, we set $\gamma{''}^{\s a}_b=\gamma{'}^{\s a}_b=0$, which implies that $g''_{ab}(x)=g'_{ab}(x)={\tilde g}_{ab}(x)$. Thus the first form of reduced reproducing kernel is given by
            \bn \<\pi''|\pi'\>\hskip-1.3em&&\equiv\<\pi'',{\tilde g}|\pi',{\tilde g}\>\no\\
                 &&=\exp(-2\tint b(x)\,d^3\!x\{\ln\det(\delta^a_b+i\s b(x)^{-1}[{\tilde\pi}{''}^{\s a}_b(x)
                 -{\tilde\pi}{'}^{\s a}_b(x)])\})\;. \en
            Note carefully: The notation $|\pi\>$ as used here does {\it not} mean a sharp eigenvector for
            the putative local operator $\hp^{ab}(x)$.
            In the second form, we choose $\pi{''}^{\s ab}(x)=\pi{'}^{\s ab}(x)=0$, which leads to the  reduced reproducing kernel given by
            \bn \<g''|g'\>\hskip-1.3em&&\equiv\<0,g''|0,g'\>\no\\
              &&=\exp(-2\tint b(x)\,d^3\!x\s \ln\{[\det(\half{\tilde g}''^{\s a}_b(x)+\half {\tilde g}'^{\s a}_b(x))]\times\no\\
              &&\hskip4em \big{/}[\det({\tilde g}''^{\s a}_b(x))\,\det({\tilde g}'^{\s a}_b(x))\s]^{1/2}\})\;.\en
             As before, the notation $|g\>$ does {\it not} refer to an eigenvector of the local operator $\hg_{ab}(x)$.

             Both of these expressions serve as reproducing kernels for the same abstract Hilbert space  that arose as the abstract Hilbert space associated with the original reproducing kernel.
             The fact that these reductions still span the original space follows from an examination of the expression
                \bn \<k|\Pi,\Gamma\>=C\s (\det[S])^2(\det[SkS^T])^{\beta-1}\,e^{i\s\tr(\Pi\s k)-\beta\s\tr({\tilde G}^{-1}S\s k\s S^T)}\;, \en
                which is the elementary building block of an affine coherent state at a single site in a lattice regularization; see (\ref{e77}). If the inner product with a general element $\psi(k)=\<k|\psi\>$, given by
                    \bn \<\psi|\Pi,\Gamma\>\equiv C \int_+\psi(k)^*\,(\det[S])^2(\det[SkS^T])^{\beta-1}\,e^{i\s\tr(\Pi\s k)-\beta\s\tr({\tilde G}^{-1}\s S\s k\s S^T)}\,dk\;,\en
               vanishes for all matrices $\Pi$ and $S\equiv\exp(-\Gamma/2)$, then it follows that $\psi(k)=0$, almost everywhere, thereby determining the span of the affine coherent states
               as $L^2_+(R^6,dk)$. Note well that the same result holds if we set either $\Gamma=0$ or $\Pi=0$, which establishes that the span of each of the  reduced reproducing
               kernels discussed above leads to the same span as the original reproducing kernel.

      \section{Enforcing the Diffeomorphism Constraints}
       Let us focus on enforcing the diffeomorphism constraints for gravity given by $\H_a(x)=-2\s\hp^b_{a\s|\s b}(x)=0$ for all $a$ and $x$, where the notation $_|$ signifies a covariant derivative based on the metric $\hg_{ab}(x)$. Neglecting temporarily the Hamiltonian constraint for gravity (see Sec.~9), we observe that enforcing the
       diffeomorphism constraint operators requires that we find states $|\psi\>_{phys}$ such that $\H_a(x)\s|\psi\>_{phys}=0$. Strictly speaking, however, those solutions may not lead to the full solution space since some nonzero eigenvectors  may not be normalizable and thus they technically do not lie in the kinematical Hilbert space ${\frak H}$. Unlike some other procedures, the projection operator method of dealing with quantum constraints \cite{PROJ} can deal with constraint operators that may have their zero in the continuum. This is not the place to offer a review of the projection operator method, and while some additional features of the projection operator method are given in Sec.~9, it suffices, for present purposes,  to let $\E$ denote a genuine projection operator ($\E^\dag=\E^2=\E$) that generates the regularized physical Hilbert space ${\frak H}_{phys}\equiv\E\s{\frak H}$; here the regularization may be symbolized by a parameter $\delta>0$ that enforces the constraints to lie within a (suitably defined) spectral window between
       $\pm\s\s\delta$. This projection operator is characterized by its affine coherent state matrix elements $\<\pi'',
       g'|\s\E\s|\pi',g'\>$, which is seen to be a continuous function of positive type. If we take the limit $\delta\ra0$ in order to enforce the constraints exactly, the
       coherent state matrix
       elements of $\E$ would possibly vanish. Although different elements may vanish at different rates, the simplest examples may be recovered by a suitable rescaling. To extract the germ carried by such matrix elements, we rescale those matrix elements, e.g., by dividing by $\<\eta|\s\E\s|\eta\>$
       before taking the limit, which leads us  to a nonzero reproducing kernel for a functional Hilbert space in which the diffeomorphism constraints are fully satisfied. In symbols, this construction is given by
       \bn \<\!\<\pi'',g''|\pi',g'\>\!\>\equiv\lim_{\delta\ra0}\frac{\<\pi'',g''|\s\E\s|\pi',g'\>}{\<\eta|\s\E\s|\eta\>}\;,\en
       yielding a continuous function of positive type, $\<\!\<\pi'',g''|\pi',g'\>\!\>$, that serves as the reproducing kernel for the Hilbert space in which the diffeomorphism constraints are fully satisfied.

       At the present time, we are unable to completely carry out the program outlined above to explicitly determine the entire Hilbert space on which the diffeomorphism constraints are satisfied. However, we can do the next best thing: we can develop functional realizations of numerous subspaces of the constrained Hilbert space.  Let us outline how we intend to generate these subspaces.

\subsection{Diffeomorphism invariant reproducing kernels}
       For convenience, we temporarily abbreviate $\<\pi'',g''|\s\E\s|\pi',g'\>$ as $\<''|\s\E\s|'\>$ letting the bra and ket refer to general affine coherent states. As  a projection operator, $\E$ admits an
       expansion---perhaps only partially---as $\Sigma_i\s|i\>\<i|$, where $\{|i\>\}$ denotes a suitable set of orthonormal vectors. Thus, the associated reproducing kernel takes the form $\Sigma_i\,\<''|i\>\<i|'\>$. As a reproducing kernel, the same space of continuous functions arises if we rescale the given vectors $|i\>$ leading to a new reproducing kernel
       $\Sigma_i\,r_i\s\<''|i\>\<i|'\>$, where $\{r_i\}$ is a set of positive real numbers; this change amounts to a similarity transformation of the previous Hilbert space representation. Moreover, to create a general reproducing kernel, the states $|i\>$ need not be mutually orthogonal but rather just linearly independent. Thus we are led to consider suitable reproducing kernels that are constructed as
          \bn \<\!\<''|'\>\!\>\equiv \Sigma_i\s c''^*\s\<''|i\>\<i|'\>\s c'\;, \en
       where we have generalized the notation so that the states $\{|i\> \}$ need only be linearly independent of one another and of arbitrary (nonzero, but finite) norm, as well as adding additional, arbitrary, nonzero coefficients $c''^*$ and $c'$, which can be used---as we do in the following examples---to cancel the
       normalizing factors in the affine coherent states.

       The next question we consider is how are we to chooses vectors $\{|i\>\}$ that exhibit diffeomorphism invariance. We first note that the expression (\ref{k22}) for the original coherent state overlap $\<\pi'',g''|\pi',g'\>$ is {\it invariant} (apart from $b(x)$) under a common coordinate transformation for {\it both} pairs of
       labels, i.e., $(\pi'',g'')$ {\it and} $(\pi',g')$. On the other hand, the idealized diffeomorphism invariant reproducing kernel $\<\!\<\pi'',g''|\pi',g'\>\!\>$ should be invariant under {\it separate} coordinate transformations of each set of the $''$ and $'$ variables; this property embodies the traditional concept that only the ``geometry of space'' carries the proper physics. How do we capture this aspect? We do so by carefully choosing suitable vectors $\{|i\>\}$
       that serve the required purpose. Ignoring the coherent state normalization factor, one conceivable example might be $\<\pi,g|1\>=\tint \sqrt{g(x)}\,d^3\!x$, where as customary we let $g(x)=\det(g_{ab}(x))$. However, this example will not work for the following reason. The affine coherent states $\{|\pi,g\>\}$ are, apart from the normalization factor, functions of the {\it complex}
       symmetric  tensor ${g}^{(-)\s ab}(x)\equiv {g}^{ab}(x)-i\s b(x)^{-1} {\pi}^{ab}(x)$. Likewise,
       the adjoint affine coherent states $\{\<\pi,g|\}$, again apart from the normalization factor, are functions
       of the complex symmetric tensor ${g}^{(+)\s ab}(x)\equiv {g}^{ab}(x)+i\s b(x)^{-1}{\pi}^{ab}(x)$. Thus, again ignoring the normalization factor in $\<\pi,g|$, as we shall continue to do, the
       diffeomorphism invariant state should rather be taken as
       \bn \<\pi,g|1\>\hskip-1.3em&&=\tint \sqrt{g^{(+)}(x)}\,d^3\!x \;.\label{k45}\en
        {\bf [Remark:} So long as $\det(g)>0$ it follows that $\det(g^{(\pm)})\not=0$, and therefore $g^{(\pm)}_{\,ab}(x)$ exists such that $g^{(\pm)}_{\,ab}(x)\s g^{(\pm)\,bc}(x)=\delta^c_a$.{\bf]} However, (\ref{k45})  is not a good state to use for noncompact spatial surfaces. Provided there is suitable asymptotic behavior, we suggest other examples worth considering, such as
        \bn    &&\<\pi,g|2\>=\tint R^{(+)}(x)\,\sqrt{g^{(+)}(x)}\,d^3\!x\;,\no\\
             &&\<\pi,g|3\>=\tint R^{(+)\,2}(x)\,\sqrt{g^{(+)}(x)}\,d^3\!x\;,\no\\\
             &&\<\pi,g|4\>=\tint R^{(+)}_{ab}(x)\s R^{(+)\,{ab}}(x)\,\sqrt{g^{(+)}(x)}\,d^3\!x\;,\\
             &&\<\pi,g|5\>=\tint R^{(+)}_{\s |\s  ab}(x)\,g^{(+)\,ab}(x)\,\sqrt{g^{(+)}(x)}\,d^3\!x\;,\no\\
             &&\<\pi,g|6\>=\tint R^{(+)\,abcd}(x)\s R^{(+)}_{\s |\s abcd}(x)\,\sqrt{g^{(+)}(x)}\,d^3\!x\;,\no\en
     etc., where $R^{(+)}(x)$ is the (three-dimensional) scalar curvature constructed from the metric $g^{(+)}_{ab}(x)$, and so on
     for the Ricci and Riemann tensors and their covariant derivatives denoted by $_|$. Clearly, each of these
     expressions is invariant under arbitrary coordinate transformations, and they would also admit extension to a noncompact space ${\cal S}$ provided the indicated integrals converged, which, for example, would be the case if ${\tilde g}_{ab}(x) = \delta_{ab}$.

     The five-dimensional
     Hilbert space that is generated by the reproducing kernel
     \bn  \<\!\<\pi'',g''|\pi',g'\>\!\>\equiv \sum_{i=2}^6 \<\pi'',g''|i\>\<i|\pi',g'\>  \en
     enjoys complete and independent invariance of both label sets under arbitrary coordinate transformations; it also provides a ``toy'' example of what is possible to generate by these techniques.

     As a further example that incorporates an infinite-dimensional Hilbert space, we offer the reproducing kernel
     \bn  &&\<\!\<\pi'',g''|\pi',g'\>\!\>\equiv \exp{\bigg\{}-\tint R''^{(+)\,2}(x)\,\sqrt{g^{''\,(+)}(x)}\,d^3\!x \no\\
         &&\hskip3em +\tint R''^{(+)\,}(x)\,\sqrt{g^{''\,(+)}(x)}\,d^3\!x \,\cdot\,\tint R'^{(-)}(x)\,\sqrt{g^{'(-)\,}(x)}\,d^3\!x\no\\
         &&\hskip3em -\tint R'^{(-)\,2}(x)\,\sqrt{g^{'(-)\,}(x)}\,d^3\!x {\bigg\}}\;. \en
     One additional example appears later in Sec.~9.1.

     \subsection{Reduction of diffeomorphism invariant\\ reproducing kernels}
     Just as we discussed the reduction of reproducing kernels for the original affine coherent state reproducing kernel, we can consider a similar reduction of those  reproducing kernels that are designed to satisfy  diffeomorphism invariance. As a particularly interesting example, we set $\pi''^{\s ab}(x)=\pi'^{\s a}(x)=0$ leading, e.g., to $\<g|1\>\equiv\<0,g|1\>=\tint \sqrt{g(x)}\,d^3\!x$, which involves a real metric. As a further example, we note that
       \bn \<g|2\>=\tint R(x)\,\sqrt{g(x)}\,d^3\!x\;, \en
       which again involves traditional geometric elements.  This same procedure can lead, for example, to
       \bn  \<\!\<g''|g'\>\!\>\equiv \sum_{i=2}^6 \<0,g''|i\>\<i|0,g'\>  \en
       These reproducing kernels do not involve the scalar density $b(x)$  and may therefore be considered
       to be preferable. On the other hand,
       analyticity permits an immediate extension of these particular reductions to restore the missing term $\pm\s \s i\s b(x)^{-1}\pi^{ab}(x)$ ensuring
       that the reduced reproducing kernel spans the same space as does that derived from the original   reproducing kernel, $\<\!\<\pi'',g''|\pi',g'\>\!\>$.
       Of course, other examples of this sort are easy to generate.

\section{Functional Integral Formulation}
The Hamiltonian constraint is related to the embedding of spatial slices nearby one another in the putative time
direction, and thus its inclusion within a single spatial slice is not evident. Recall that the Hamiltonian
is wholly specified at a single moment of time, but, along with other elements, it can be extended into
the time domain by the introduction of a path integral. A similar procedure is available to us.

The analyticity of the arguments of the affine coherent states up to normalization factors, which just was made use of in proposing some diffeomorphism invariant examples of Hilbert spaces, can be put to good use in a rather different manner to build a path integral
representation of the affine coherent state overlap itself \cite{GUTZ,HART}. The success of this
procedure stems from the fact that every affine coherent state representative, $\psi(\pi,g)\equiv\<\pi,g|\psi\>$, satisfies a
 {\it complex polarization condition}, namely
\bn&&\hskip-.5cm C^r_s(x)\,\psi(\pi,g)\equiv\bigg[-ig^{rt}(x)\frac{\delta}{\delta\pi^{ts}(x)}+\delta^r_s+b(x)^{-1}g_{st}(x)\frac{\delta}{\delta g_{tr}(x)}\bigg]\,\psi(\pi,g)=0 \no\\
&&\en
for all spatial points $x$ and any function $\psi(\pi,g)\in{\cal C}$, which is the associated reproducing kernel Hilbert space.
   This is a first-order functional differential equation because we have chosen the fiducial vector as an
   extremal weight vector which is determined as the solution of a linear equation in Lie algebra generators.  Multiplication by the adjoint of that first-order differential operator, followed by summation over indices and integrated over space, leads to a nonnegative, second-order functional differential operator
   given by
  \bn A\equiv \half\int C^s_r(x)^\dagger\,C^r_s(x)\,b(x)\,d^3\!x\;, \en
   with the property that $A\ge0$ , and which
    annihilates every affine coherent-state function $\psi(\pi,g)$.
    Thus, with $T>0$ and as $\nu\ra\infty$, it follows that $e^{-\nu T A}$
formally becomes a projection operator onto the space $\cal C$.
   It is clear that the second-order functional differential operator $A$ is an analog of a Laplacian operator in the presence of a magnetic field, and thus
a Feynman-Kac-Stratonovich path (i.e., functional) integral representation may be introduced. In particular, we are led to the formal expression
  \bn&&\hskip-1cm\<\pi'',g''|\pi',g'\>=\lim_{\nu\ra\infty}{\cal N}\int\exp[-i\tint g_{ab}{\dot\pi}^{ab}\,d^3\!x\,dt]\no\\
 &&\hskip.3cm\times\exp\{-(1/2\nu)\tint[b(x)^{-1}g_{ab}g_{cd}{\dot\pi}^{bc}{\dot\pi}^{da}+b(x)g^{ab}g^{cd}{\dot g}_{bc}{\dot g}_{da}]\,d^3\!x\,dt\}\no\\
&&\hskip4cm\times {\t\prod}_{x,t}{\t\prod}_{a\le b}\,d\pi^{ab}(x,t)\,dg_{ab}(x,t)\;.  \label{ff3}
\en
It this expression, it is natural to interpret $t$, $0\le t\le T$,  as coordinate ``time'', and thus on the right-hand side the canonical fields are functions of space and time, that is
  \bn g_{ab}=g_{ab}(x,t)\;,\hskip1cm \pi^{ab}=\pi^{ab}(x,t)\;,  \en
where the overdot
($\,{\dot {~}}\,$)
denotes a partial derivative with respect to $t$, and the integration is subject to the boundary conditions that $\pi(x,0),\,g(x,0)=\pi'(x),\,g'(x)$ and $\pi(x,T),\,g(x,T)=\pi''(x),\,g''(x)$.
It is important to note, for any $\nu<\infty$, that underlying the formal expression (\ref{ff3}) given above, there is a genuine, countably additive measure on (generalized) functions $g_{kl}$ and $\pi^{rs}$. Loosely speaking, such functions have Wiener-like behavior with respect to time and $\delta$-correlated, generalized Poisson-like behavior with respect to space.

It is important to understand that although the functional integral (\ref{ff3}) is over the canonical momentum  $\pi^{ab}(x,t)$ and the canonical metric  $g_{ab}(x,t)$, this integral has arisen strictly from an affine quantization and {\it not} from a canonical quantization. It also noteworthy that the very structure of the $\nu$-dependent regularization factor forces the metric to satisfy metric positivity, i.e., $u^a\s g_{ab}(x,t)\s u^b>0$, provided that $\Sigma_a (u^a)^2>0$.

Phase space path integrals with Wiener-measure regularization were introduced in \cite{DAUB}. The regularization involves a phase space metric, and on examining (\ref{ff3}) it is clear that the phase space metric, as given, is almost uniquely determined.

\section{Imposition of All Constraints}
Gravity has four constraints at every point $x\in{\cal S}$, and, when expressed in suitable units, they are the familiar spatial (diffeomorphism) and temporal (Hamiltonian) constraints, all densities of weight one, given by \cite{ADM,mtw}
 \bn && H_a(x)=-2\s\pi^{b}_{a\;|\s b}(x)\;, \no \\
&&H(x)= g(x)^{-1/2}[\s\pi^a_b(x)\s\pi^b_a(x)-\half\s\pi^c_c(x)\s\pi^d_d(x)\s]+g(x)^{1/2}\,R(x)\;, \en
where, as has been our custom, $R(x)$ is the three-dimensional scalar curvature.
While spatial constraints are comparatively easy to incorporate as we have seen in Sec.~7,  this is  not the case for the temporal constraint. A detailed account of how all the constraints can be accommodated in constructing a regularized projection operator $\E^*$ for them has already been given in
 \cite{AQG2} and need not be repeated here. The notation $\E^*$ is meant to distinguish the strictly diffeomorphism constraint projection operator $\E$ from the all-constraint projection operator $\E^*$.

Briefly stated, the introduction of the all-constraint projection operator begins with the continuous-time regularized functional integral representation of the affine coherent state
reproducing kernel $\<\pi'',g''|\pi',g'\>$ for the kinematical Hilbert space. The reproducing kernel for the regularized physical Hilbert space is given, in turn, by the expression $\<\pi'',g''|\s\E^*\s|\pi',g'\>$,
where $\E^*$ refers to a projection operator that includes all constraint operators. In order
to give this latter expression a functional integral representation we initially assume that
   \bn  \tint[N^a\s\H_a+N\s\H]\,d^3\!x  \en
is a time-dependent ``Hamiltonian'' for some fictitious theory, in which case
 \bn  && \<\pi'',g''|\s{\bf T}\,e^{\s-i\s\tint[N^a\s\H_a+N\s\H]\,d^3\!x\,dt}\s|\pi',g'\> \no\\
  &&\hskip1cm=\lim_{\nu\ra\infty}{\o{\cal N}}_\nu\,\int\exp\{-i\tint[g_{ab}\s{\dot\pi}^{ab}+N^a\s H_a+N\s H]\,d^3\!x\,dt\} \no\\
  &&\hskip1.5cm\times\exp\{-(1/2\nu)\tint[b(x)^{-1}g_{ab}g_{cd}{\dot\pi}^{bc}{\dot\pi}^{da}+b(x)g^{ab}g^{cd}{\dot g}_{bc}{\dot g}_{da}]\,d^3\!x\,dt\}\no\\
  &&\hskip2cm\times\Pi_{x,t}\,\Pi_{a\le b}\,d\pi^{ab}(x,t)\,dg_{ab}(x,t)\;. \label{f35} \en
In this expression, there appear symbols $H_a(\pi,g)$ and $H(\pi,g)$ corresponding to the quantum operators $\H_a$ and $\H$.  We do not discuss the details of these symbols, which are unlikely to simply be the classical constraint expressions due to the fact that $\hbar>0$ within this functional integral.

To pass from the intermediate stage of our fictitious theory to the final formulation involving the regularized projection operator, we introduce an integration over the variables $N^a$ and $N$ using a
carefully constructed measure ${\cal D}\s R(N^a,N)$, which is defined in \cite{AQG2}. This leads to the expression
 \bn  && \<\pi'',g''|\s\E^*\s|\pi',g'\> \no\\
  &&\hskip1cm=\lim_{\nu\ra\infty}{\o{\cal N}}_\nu\,\int\exp\{-i\tint[g_{ab}\s{\dot\pi}^{ab}+N^a\s H_a+N\s H]\,d^3\!x\,dt\} \no\\
  &&\hskip1.5cm\times\exp\{-(1/2\nu)\tint[b(x)^{-1}g_{ab}g_{cd}{\dot\pi}^{bc}{\dot\pi}^{da}+b(x)g^{ab}g^{cd}{\dot g}_{bc}{\dot g}_{da}]\,d^3\!x\,dt\}\no\\
  &&\hskip2cm\times[\Pi_{x,t}\,\Pi_{a\le b}\,d\pi^{ab}(x,t)\,dg_{ab}(x,t)]\,{\cal D}R(N^a,N)\;. \label{f39} \en
  The result of this functional integral, $\<\pi'',g''|\s\E^*\s|\pi',g'\>$, is a continuous function of positive type that can be used as a reproducing kernel for a Hilbert space in which the full set of gravitational constraints are satisfied in a regularized manner. {\bf[Remark:} Incidentally, omitting the integral over $N$ and setting $H=0$ altogether would lead to the result $\<\pi'',g''|\s\E\s|\pi',g'\>$, which, as noted previously, is a continuous function of positive type that would serve as a reproducing kernel for the Hilbert space in which only the diffeomorphism constraints are satisfied in a regularized manner.{\bf]}

  Unfortunately, the  evaluation of such functional integrals is beyond present capabilities. However, just
   as we were able to introduce sub spaces that satisfy all the diffeomorphism constraints alone, there should exist similar examples of sub spaces where all of the constraints are satisfied, and which should  have a form not unlike that for the diffeomorphism constraint situation.
   In fact, in the next
   subsection we argue that the examples already developed for the diffeomorphism constraints should also work for the case of all the quantum gravity constraints.

   \subsection{A proposal for reproducing kernels satisfying all constraints}
   The affine coherent state matrix elements of the two projection operators, $\E$ and $\E^*$, have much
   in common. Apart from affine coherent state normalization factors, it follows that both $\<\pi'',g''|\s\E\s|\pi',g'\>$ and $\<\pi'',g''|\s\E^*\s|\pi',g'\>$ are functions of $g^{(\pm)\,ab}(x)$. Both functions have to depend on strictly geometric combinations such as $R^{(\pm)}(x)$, etc., so as to be invariant under any coordinate transformation. Moreover, the original coherent states were defined on a space-like surface in the topological carrier space ${\cal S}$, but no particular space-like surface has been specified. In other words, the coherent states and their matrix elements of various operators of interest such as $\E$ and $\E^*$, are invariant under arbitrary changes of the space-like surface on which the coherent states have been defined.

   It is noteworthy that fulfillment of the diffeomorphism constraints on all space-like surfaces nearly implies that  the Hamiltonian constraint is fulfilled as well, according to the following argument kindly supplied by Karel Kucha$\check{\rm r}$. Let $G_{\mu\nu}(x)$ denote the usual Einstein tensor, and the general fulfillment of the diffeomorphism constraints is given by the relation $n^\mu_{\bot}(x)\s
   \sigma^\nu_\bot(x)\s G_{\mu\nu}(x)=0$ for a general time-like vector $n^\mu_\bot(x)$ and a general space-like vector $\sigma^\nu_\bot(x)$, which, however, are required to fulfill $n^\mu_\bot(x)\s\sigma^\nu_\bot(x)\s g_{\mu\nu}(x)\equiv 0$ (hence the notation with $\bot$). The perpendicularity relation can be accommodated by  Lagrange multipliers, leading to the relation $n^\mu(x)\s \sigma^\nu(x)\s[\s G_{\mu\nu}(x)+\l(x)\s g_{\mu\nu}(x)]=0$,
   now with no restriction that $n^\mu(x)$ and $\sigma^\nu(x)$ are explicitly perpendicular. Since the difference of two time-like vectors may be space like, and the difference of two space-like vectors may be time like, we conclude that $G_{\mu\nu}(x)+\l(x)\s g_{\mu\nu}(x)=0$. However, a contracted Bianchi identity implies that
   $G^\mu_{\nu\s\s;\mu}(x)=0=-[\l(x)\s\delta^\mu_\nu]_{\s\s:\mu} \s$, or finally that $\l(x)=\Lambda$, a constant;
    indeed, if the space is asymptotically flat, then the fact that $\lim_{|x|\ra\infty} G_{\mu,\nu}(x)=0$ implies that $\Lambda=0$.
   In summary, then, the fulfillment of the diffeomorphism constraints on general space-like surfaces implies that the Hamiltonian constraint is also satisfied, up to a generally unspecified cosmological constant.
   Consequently, we are led to propose that the various diffeomorphism invariant reproducing kernel examples given in Sec.~7 are valid as well for the full set of quantum constraints.

   To give one more example, we offer another infinite-dimensional reproducing kernel given by
     \bn \<\!\<\pi'',g''|\pi',g'\>\!\>=\int\!\!\int\s\{\exp[\s R^{(+)}(x)\s R^{(-)}(y)]-1\}\,\sqrt{g^{(+)}(x)\,g^{(-)}(y)}\,d^3\!x\s d^3\!y. \en
     Observe, that in this expression the two integrals should be over the same space-like surface, but that surface is an arbitrary space-like surface. This reproducing kernel, as well as the other examples in Sec.~7, may be considered as ``toy'' examples each one being given, effectively, by its own projection operator $\E_T$ ($T$ for toy), which, in a picturesque sense, satisfies,
     $\E_T\subset\E$. In the same sense, $\E^*\subset\E$, and it is not out of the question that suitable toy models may be related as $\E_T\subset \E^*\subset \E$, and thus those toy examples would fulfill all of the quantum gravity constraints.

   \section{Conclusions}
   The program of affine quantum gravity is very conservative in its approach. It adopts as
   classical variables the spatial momentric $\pi^a_b(x)\equiv \pi^{ac}(x)\s g_{bc}(x)$ and the spatial metric $g_{ab}(x)$, and taken together this set of kinematical variables is called the affine variables. {\bf[Remark} Note that the traditional momentum may readily be found from the momentric by $\pi^{ab}(x)=\pi^a_c(x)\s g^{bc}(x)$.{\bf]}
   The focus on the affine variables arises from the effort to ensure that the chosen kinematical variables universally respect metric positivity, namely, $u^a\s g_{ab}(x)\s u^b>0$, provided $\Sigma_a (u^a)^2>0$. Upon quantization, as well, the affine variables, which satisfy the affine commutation relations, respect metric positivity, unlike the canonical variables which obey the usual canonical commutation relations.
   The affine commutation relations are like current commutation relations, and as such they admit representations that are bilinear in conventional creation and annihilation operators, which are quite unlike the representations generated by canonical commutation relations. In particular, the bilinear realizations of the affine variables admit straightforward local products by operator product expansion constructions that do not involve normal ordering of any kind. Consequently, just as the spatial metric operator $\hg_{ab}(x)$ becomes a self-adjoint operator when smeared by a real test function, this same property
   holds for the local product $[\s g_{ab}(x)\s g_{cd}(x)\s]_R$, and all other operator-product renormalized local operators. Therefore, it is clear, e.g., that the sum of such expressions integrated over a finite volume generates a self-adjoint operator, which might be part of the Hamiltonian. Stated otherwise, the class of representations consistent with an operator product expansion offers profoundly greater
   opportunities to construct various local operators  of interest that are nonlinear functions of  the basic kinematical operators. Going hand-in-hand with the bilinear operator representations, is the realization of such operators as well as their local products within one and the same Hilbert space, meaning that one and only one irreducible representation is involved due to the implicit measure mashing that occurs for such representations. While measure mashing is wholly foreign to operators that obey canonical commutation relations, it is---and at first glance surprisingly so---already built into the representation structure of the affine operators. A theory that involves measure mashing leads to functional integral representations which, under changes of its parameters as part of a perturbation analysis, entail different measures that are {\it equivalent} (equal support) rather than being (at least partially) {\it mutually singular} (disjoint support), and for that very reason, do not create term-by-term divergences; to observe measure mashing in action see \cite{IOP}. The importance of measure mashing, and its automatic incorporation by the affine quantum gravity kinematic variables, is a new insight for the author since the
   appearance of the principal papers on affine quantum gravity \cite{AQG1,AQG2}.

   Another realization that is new in the present paper is the ability to create a vast number of reproducing kernels---and implicitly thereby the Hilbert space representations by continuous functions they create---that are fully invariant under the action of the diffeomorphism constraints. Moreover, the labels of the various reproducing kernels are the familiar classical variables of the canonical theory, namely, a smooth tensor density of weight one, the spatial momentum $\pi^{ab}(x)$. and a smooth spatial metric tensor $g_{ab}(x)$, restricted as always by metric positivity. One can imagine that these subspaces, which can have infinite dimensionality, may be useful in the study of suitable toy models.

   The functional integral representation for: (i) the overlap function of the affine coherent states,
   (ii) the affine coherent state matrix elements of the projection operator enforcing just the diffeomorphism constraints, and (iii) the affine coherent state matrix elements of the projection operator enforcing all
   of the constraints involves a common language within which these three distinct sets of matrix elements can be considered. In particular, the coherent state overlap function clearly illustrates the functional dependence of the coherent states on their labels. This functional dependence also holds for the matrix elements of the
   two separate projection operators of interest, namely $\E$ and $\E^*$. For the diffeomorphism constraints alone, we were able to make logical choices for the functional dependence of suitable sub spaces that remain consistent with the requirements involved in the coherent state overlap function itself.

   It is tempting to believe that there is yet another set of proper arguments that would allow us to set up sub spaces of the reproducing kernel in which all of the constraints are fulfilled. We have proposed that the several examples of sub spaces that satisfy the diffeomorphism constraints also satisfy all the quantum constraints, but that hypothesis needs to be carefully examined further to see if it carries any truth.

   \section*{Acknowledgements}
   Thanks are offered to K.V.~Kucha$\check{\rm r}$, G.~Watson, and B.~Whiting for helpful contributions.

\end{document}